\documentclass[useAMS,usenatbib,twocolumn,preprintnumbers,nofootinbib]{revtex4}
\usepackage{graphicx}% Include figure files
\usepackage{dcolumn}% Align table columns on decimal point
\usepackage{bm}% bold math

\begin{document}

\preprint{GCON-2015/07}

\title{Baryon Acoustic Oscillations from the SDSS DR10 galaxies angular correlation function}

\author{G. C. Carvalho}
\email{gabriela@on.br}
\author{A. Bernui}
\email{bernui@on.br}
\author{M. Benetti}
\email{micolbenetti@on.br}
\author{J. C. Carvalho}
\email{jcarvalho@on.br}
\author{J. S. Alcaniz}
 \email{alcaniz@on.br}
\affiliation{Observat\'orio Nacional, 20921-400, Rio de Janeiro, RJ, Brazil}

\date{\today}

\begin{abstract}

The 2-point angular correlation function $w(\theta)$ (2PACF), where $\theta$ is the angular separation between pairs of galaxies, provides the transversal Baryon Acoustic Oscillation (BAO) signal almost model-independently. In this paper we use 409,337 luminous red galaxies in the redshift range $z = [0.440,0.555]$ obtained from the tenth data release of the Sloan Digital Sky Survey (SDSS DR10) to estimate $\theta_{\rm{BAO}}(z)$ from the 2PACF  at six redshift {shells}. Since noise and systematics can hide the BAO signature in the $w - \theta$ plane, we also discuss some criteria to localize the acoustic bump. We identify two sources of model-dependence in the analysis, namely, the value of the acoustic scale from Cosmic Microwave Background (CMB) measurements and the correction in the $\theta_{\rm{BAO}}(z)$ position due to projection effects. Constraints on the dark energy equation-of-state parameter w$(z)$  from the  $\theta_{\rm{BAO}}(z)$ diagram are derived, as well as from a joint analysis with current 
CMB measurements. We find that the standard $\Lambda$CDM model as well as some of its extensions are in good agreement with these $\theta_{\rm{BAO}}(z)$ measurements.

\end{abstract}

% PACS, the Physics and Astronomy 
% Classification Scheme.

\maketitle

\section{\label{sec:level1}Introduction}

Baryon Acoustic Oscillations (BAO) arise due to the competing effects of radiation pressure and gravity in the early Universe~\cite{PeeblesY,SZ,BE,Hu-Dodelson,BG}. At $z \sim 1000$ photons and baryons decouple and a characteristic scale $r_s$, corresponding to the sound horizon at the drag epoch, is imprinted in the maps of galaxy distribution and in the power spectrum of cosmic microwave background anisotropies\footnote{The acoustic scale is defined as $r_s = \int_{z_d}^\infty{c_s(z) \over H(z)}dz$, where $z_d \approx 1070$ is the redshift of the drag epoch and $c_s(z)$ is the sound speed of the photon-baryon fluid.}. In recent analysis, a tiny excess of probability to find pairs of galaxies separated by a distance equal to this comoving acoustic radius was revealed in the 2-point correlation function (2PCF) of galaxy catalogs, where it appears as a bump. The first detections of this BAO signature were obtained from galaxy clustering analysis of the Two Degree Field Galaxy Survey (2dFGRS)~\cite{cole05} and 
from 
the Luminous Red Galaxies (LRGs) data of the Sloan Digital Sky Survey (SDSS)~\cite{Eisenstein05}. A subsequent joint analysis of 2dFGRS and SDSS data yielded a BAO distance measurement with aggregate precision of 2.7\% at redshift $z=0.275$~\cite{Percival10} (see also \cite{pad}). More recently, higher-$z$ measurements at percent-level precision were also obtained using deeper and larger galaxy surveys \cite{Percival10,Beutler11,Blake11b,Sanchez12} (see also \cite{eisensteinreview} for a recent review).

The BAO signature defines a robust {standard ruler}, providing independent measures of the angular diameter distance $D_{\!A}(z)$ and the Hubble parameter $H(z)$ through the transversal and radial BAO modes, respectively. However, the detection of the BAO signal through the 2PCF makes necessary the introduction of a {fiducial cosmology} to calculate the radial distance to the cosmic objects then the comoving distance between them.  On the other hand, the calculation of the 2-point {{angular}} correlation function (2PACF) involves only the angular separation between pairs, yielding model-independent information about $\theta_{\rm{BAO}}(z)$ or, equivalently, $D_{A}(z)$, provided that robust estimates of the comoving acoustic scale are obtained. Therefore, determining the angular position of the BAO bump at several narrow redshift shells allows {us} to test the observational viability of different dark energy models through 
the $\theta_{\rm BAO} - z$ diagram. In this type of analysis, narrow redshift {shell} are necessary in order to avoid contributions from the radial BAO signal. 

In recent years, efforts have been made to map increasingly large volumes of the sky and to explore the cosmological consequences of the BAO signature imprinted in galaxies distribution. One of the major multi-filter imaging and spectroscopic redshift surveys is the Sloan Digital Sky Survey (SDSS) that has been operating over fourteen years~\cite{SDSS}. In this work we use the tenth public data release (DR10) \cite{Anderson14} of BOSS experiment, part of the SDSS-III project, to perform {\emph{almost}} {model-independent} cosmological analyses  using the BAO transversal signature obtained from the 2PACF. We use only 2-dimensional information, which restricts us to focus on the 2PACF in a set of narrow redshift {shells}. As the 2PACF is usually noisy due to  systematic effects, we also introduce a double-tool methodology to identify the BAO bump. Our almost model-independent approach is possible due to some characteristics of the SDSS DR10 which were absent in previous releases. As will be clear in the next 
section, the most important one is the 
high galaxy number density which allows it to have sufficient galaxy-correlated pairs revealing the angular BAO signature in small redshifts {shells} ($\delta z = 0.01$ and 0.02).

This paper is organised as follows. In Sec. II  we describe the basic methodology of our study, including the relevant underlying equations for the 2PACF analyses. Sec. III discusses the observational data employed in this paper and the criteria for selecting the redshift shells. A new methodology to identify the BAO signal from the 2PACF curves as well as the underlying cosmological assumptions to extract $\theta_{\rm{BAO}}(z)$ from the LRGs catalogs are described in Sec. IV. Assuming a time-dependent parameterisation for the dark energy equation of state, in Sec. V we discuss the cosmological constraints on the parameters of dark energy models from the $\theta_{\rm{BAO}}(z)$ diagram. Our overall conclusions are summarised in Sec. VI.

\section{\label{sec:level1}The 2-point angular correlation function} 

The {two-point correlation function} (2PCF), $\xi(s)$, $s$ being the comoving separation, is usually used to determine the BAO feature given a set of cosmic tracers like, e.g., Luminous Red Galaxies (LRGs). This function is defined as the excess probability of finding two pairs of galaxies at {a} given distance and is  obtained by comparing the real catalog to random catalogs  that follow the geometry of the survey.  Among several estimators of the 2PCF discussed in the literature (see, e.g.,~\cite{Peebles-Hauser, Hewett,Davis-Peebles,Hamilton,Landy-Szalay}), one of the most commonly used is the one proposed in Ref.~\cite{Landy-Szalay} (and adopted in our analyses):
\begin{equation}
\xi (s) = \frac{DD(s) - 2 DR(s) + RR(s)}{RR(s)} \, . 
\end{equation}
{Here,} $DD(s)$ and $RR(s)$ correspond to the number of galaxy pairs with separation $s$ in real-real and random-random catalogs, respectively, whereas $DR(s)$ stands for the number of pairs with comoving separation $s$ calculated between a real-galaxy and a random-galaxy. 
%catalog. 
As mentioned earlier, the comoving distance $s$ between pairs of galaxies is calculated assuming a {fiducial} cosmology. In fact, assuming a flat universe, as indicated by recent CMB data~\cite{wmap9,planck}, the comoving distance $s$ between a pair of galaxies at redshifts $z_1$ and $z_2$ is given by 
\begin{equation}
s = \sqrt{r^2(z_1) \,+\, r^2(z_2) \,-\, 2 \,r(z_1) \,r(z_2)  \cos \theta_{12} \,\,} \, ,
\end{equation}
where $\theta_{12}$ is the angular distance between such a pair of galaxies, and the radial distance between the observer and a galaxy at redshift $z_i$, $r(z_i)$, depends on the  cosmological model adopted. For instance, for a flat dark energy model with equation-of-state parameter w$(z)$, it reads 
\begin{equation} \label{rz}
r(z_i) = \frac{c}{H_0} \int_{0}^{z_i} \, \frac{dz}{\sqrt{\Omega_m (1+z)^3 \,+\, 
(1-\Omega_m) \Xi(z) }}  , 
\end{equation}
where 
$$
\Xi(z) = \exp\left[3\int_{0}^{z}[1+{\rm{w}}(z')d\ln{(1+z')}]\right],
$$
$H_0$ and $\Omega_m$ are, respectively, the Hubble constant and the present-day matter density parameter. 

Analogously to the 2PCF, the 2PACF is defined as the excess joint probability that two point sources are found in two solid angle elements $d\Omega_1$ and $d\Omega_2$ with angular separation $\theta$ compared to a homogeneous Poisson distribution~\cite{peebles}. This function can be used model-independently, considering only angular separations in narrow redshift shells of small $\delta z$ in order to avoid contributions from the BAO mode along the line of sight (radial signal). The 2PACF is now calculated as a function of the angular separation $\theta$ between pairs {at a given redshift shell}, i.e.,
\begin{equation}
w(\theta) = 
\frac{DD(\theta) - 2 DR(\theta) + RR(\theta)}{RR(\theta)} \, , 
\end{equation}
where the transversal signal of the acoustic BAO scale manifests itself as a bump at certain angular scale $\theta_{FIT}$. Naturally, once the position of such BAO bump is localised, it is possible to build the $\theta_{\rm{BAO}}(z)$ diagram, i.e., the evolution of the angular diameter-distance with redshift (see Sec. VI).

One can obtain the expected 2PACF, $w_{E}$, in terms of the expected 2PCF, $\xi_{E}$, considering a distribution of objects between redshifts $z_1$ and $z_2$ as a function of the angular distance $\theta$ between pairs~\cite{Carnero,Salazar}, 
namely 
\begin{equation} \label{expected}
w_{E}(\theta, \bar{z}) = \int_0^\infty dz_1 \,\phi(z_1) \int_0^\infty dz_2 \,\phi(z_2) \,\xi_{E}(s, \bar{z}) \, , 
\end{equation}
where $\bar{z} \equiv (z_1 + z_2) / 2$, with $z_2 = z_1 + \delta z$, and  $\phi(z_i)$ is the normalised galaxy selection function at redshift $z_i$.  However, if bin shells  are narrow, $\delta z \approx 0$, then $z_1 \approx z_2$ and $\xi_{E}(s, z_1) \simeq \xi_{E}(s, z_2)$. Therefore, one can safely consider that $\xi_{E}(s, \bar{z})$ depends only on the constant parameter $\bar{z}$, instead of on the variable $z$  (for shells of arbitrary $\delta z$, see~\cite{MSP}). 
The function $\xi_{E}(s, z)$ is given by 
\begin{equation}\label{xi_e}
\xi_{E}(s,z)=\! \int_0^\infty \! \frac{dk}{2\pi^2} \, k^2 \, j_0(k s) \, b^2 \, P_m(k, z) \, , 
\label{eq:xiexp}
\end{equation}
where $j_0$ is the zeroth order Bessel function,  $P_m(k, z)$ is the matter power spectrum and the bias factor has been set at $b = 1$ since it does not affect the {BAO peak position} (for a broader discussion on the plausibility of the bias assumption and non-linear corrections, we refer the reader to \cite{Sanchez11} and references therein).

\section{\label{sec:level1}The Data set}

The phase III of the Sloan Digital Sky Survey~\cite{SDSS}, SDSS-III, had several goals which include, among others, a better understanding of the mechanism behind cosmic acceleration from the analysis of the BAO {feature} in the galaxy clustering. Four experiments have been operating to provide the DR10 \cite{Anderson14} data sets: the Sloan Extension for Galactic Understanding and Exploration (SEGUE-2), the Baryon Oscillation Spectroscopic Survey (BOSS), Apache Point Observatory Galactic Evolution (APOGEE) and Multi-Object APO Radial Velocity Exoplanet Large-area Survey (MARVELS). 

The BOSS experiment, as part of the SDSS-III project, mapped about $1.5 \times 10^6$ luminous red galaxies (LRG), as faint as $i=19.9$, over 10,000 deg$^2$, up to redshifts $z < 0.7$.  BOSS used upgraded instruments to extend the redshift survey and map deeper than the SDSS-I and II. This new spectroscopic experiment allows to map galaxy density with approximately $n = 0.0002 - 0.0003\;\mbox{h}^3 /\mbox{Mpc}^3$. 

The SDSS DR10 contains 409,337 LRGs in the north galactic hemisphere, with redshifts from $z=0.43$ to $z=0.7$, including new BOSS spectra. In order to detect transversal signatures as a function of redshift, we divided the data into six shells of redshift, as seen in Table~\ref{tab1}. The selection of the width of the redshift shell is not an easy task: too narrow redshift shells may not contain enough galaxy correlated pairs to reveal the BAO signal, whereas if the redshift shell is too wide it decreases the angular BAO amplitude, mixing the transversal signature with the radial contribution. After several analyses, we selected shells at: $\bar{z} = 0.45, 0.47, 0.49, 0.51, 0.53$, and 0.55, which are separated by a redshift interval of 0.005 to avoid correlations between them. Details of these redshift shells are presented in Table~\ref{tab1}.

\begin{figure*}[t]
\includegraphics[width=16.cm,height=11cm]{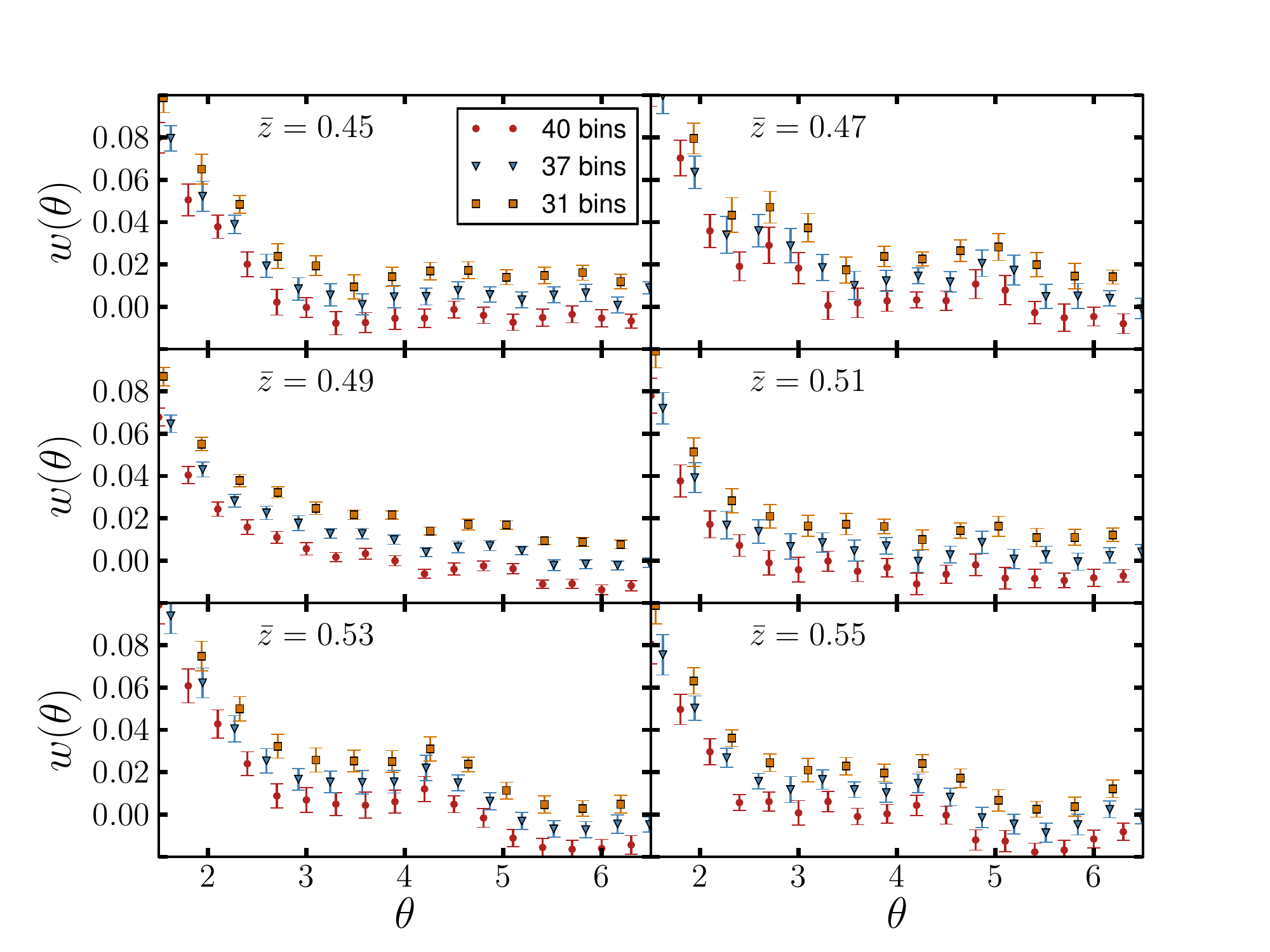}
\caption{\small Each 2PACF plot assumes different redshift {shells} with almost 20,000 galaxies 
(see details in table~\ref{tab1}). {The different $\theta$ bin corresponds to the interval from 0 to 12 degrees divided by $N_b=$ 31, 37, and 40 bins, i.e., $\Delta \theta = 12.0/N_b$.}
To a better visualization of the 2PACF curves they were artificially shifted upwards. The error bars corresponds to one standard deviation obtained from the 50 random catalogs used in the analysis.}
\label{fig1}
\end{figure*}

\begin{table}
\begin{tabular}{|  l | c | c | c | c |} 
\hline
%$\langle z \rangle$
\,\,\, redshift intervals \,\, & \,\, number of LRGs \,\, & \,\,$\bar{z}$\,\,\, 
& \,\,\,$\delta z$ \,\,\,   \\
\hline \hline
\,\,\,\, 0.440 - 0.460 \,\,\,\, & 21,862  &  \,\,\,\,0.45\,\,\,\, & \,\,\,\,0.02\,\,\,\,  \\
\,\,\,\, 0.465 - 0.475 \,\,\,\, & 17,536  &  \,\,\,\,0.47\,\,\,\, & \,\,\,\,0.01\,\,\,\,  \\
\,\,\,\, 0.480 - 0.500 \,\,\,\, & 40,957  &  \,\,\,\,0.49\,\,\,\, & \,\,\,\,0.02\,\,\,\,  \\
\,\,\,\, 0.505 - 0.515 \,\,\,\, & 21,046  &  \,\,\,\,0.51\,\,\,\, & \,\,\,\,0.01\,\,\,\,  \\
\,\,\,\, 0.525 - 0.535 \,\,\,\, & 22,147  &  \,\,\,\,0.53\,\,\,\, & \,\,\,\,0.01\,\,\,\,  \\
\,\,\,\, 0.545 - 0.555 \,\,\,\, & 21,048  &  \,\,\,\,0.55\,\,\,\, & \,\,\,\,0.01\,\,\,\,  \\
 \hline
\end{tabular}
\caption{The six bin-redshift intervals and their properties: 
number of galaxies, mean redshift of the sample, $\bar{z}$, and bin-width, $\delta z$. 
Notice that contiguous intervals are separated by a redshift interval of size 0.005 to avoid 
correlation between neighbours.} 
\label{tab1}
\end{table}

\begin{figure*}[t]
\label{fig2}
\includegraphics[width=16.cm,height=11.5cm]{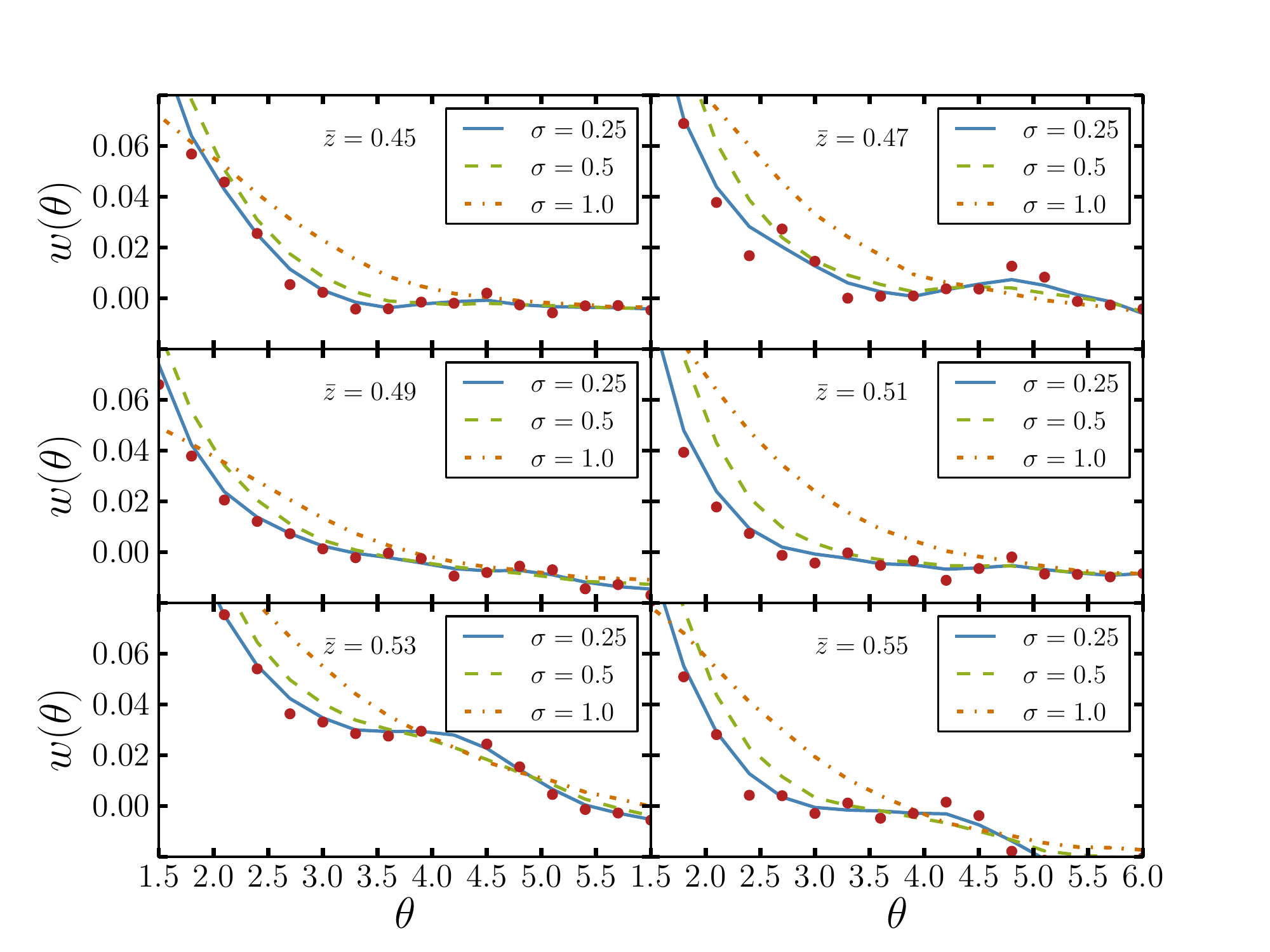}
\caption{The continuous, dashed, and dot-dashed lines correspond to the 2PACF after changing the 
angular position of galaxies by a random amount following Gaussian distributions with $\sigma=0.25, 0.5$, and 
$1.0$, respectively, whereas the dots stand for the original data. In these plots we used $N_b = 40$.} 
\label{fig2}
\end{figure*}

%%%%%%%%%%%%%%%%%%%%%%%%%%%%%%%%%%%%%%%%%%%%%%%
\section{\label{methods}Detecting the BAO signal}
%%%%%%%%%%%%%%%%%%%%%%%%%%%%%%%%%%%%%%%%%%%%%%%

In previous works (see, e.g.,~\cite{Sanchez11,Carnero}), the 2PACF was applied to galaxy surveys with photometric redshift (photo-$z$) data, where the redshift error depends on the range of  wavelengths given by {the} filters. Unfortunately, {the large error of the photometric redshift implies large uncertainties in the separations of the galaxies pairs}. In addition to this problem, some analyses need a fiducial cosmology to derive cosmological parameters from the 2PACF~\cite{deSimoni}. This strategy is usually adopted because it is very common to observe more than {one} single {bump} in any analysis of the 2PCF or 2PACF,  due to systematic effects present in the sample. Thus, when the 2PACF shows more than one bump, one needs guiding principles to recognise the true acoustic scale. Since in our case we want to perform an analysis as model-independent as possible, we identify the angular-BAO signal only when a bump remains after applying the following double-tool methodology: 

\begin{enumerate}

\item {\textit{Bin size criterium}} -- Angular separations between pairs of galaxies are counted in bin intervals with bin width $\Delta \theta$. 
Our analyses are performed for various values of $\Delta \theta$ in order to confirm if the BAO bump candidate persists or disappears. 

%There isn't a prior way to choose a ideal interval $\theta_i< \theta < \theta_i +\Delta \theta$, where $\theta_i$ is the beginning of interval.  Thus our 2PACF analyses are performed with different bin widths $\Delta \theta$.

\item {\it Small shifts criterium} -- We perform the 2PACF analysis changing the galaxies angular coordinates by small and random amount. Thus, the curve is smoothed and the bumps produced by systematic effects is removed.

\end{enumerate}

Our strategy is the following: if the BAO bump is present and robust, then it will remain after applying these criteria. If this happens, one can safely consider that the bump corresponds to a transversal BAO signature. After that, the BAO bump is {localised} through a best-fit procedure, obtaining the $\theta_{FIT}$ value, which is then corrected to the $\;\theta_{BAO}$ value  (corresponding to the ideal case in which $\delta z = 0$) using a shift correction function discussed in Sec. V.   

\subsubsection{Bin size criterium} 

In any histogram, like the 2PACF, the choice of the bin size is a compromise between a noisy curve (when narrow bins are used), where possible signatures are hidden by statistical noise and systematics, and a smooth curve (when wider bins are used), where possible signatures spread out and are almost invisible. Our leading criterion to decide the presence of a robust bump -- like a BAO bump -- is that it should persist in the 2PACF even when one changes the bin width, while noise bumps shall smear out or simply disappear. 

To optimize the choice of the bin width in our 2PACF analyses we consider several possibilities. Among them, we find three interesting cases for the angular interval, i.e., $\theta \in [0^{\circ}, 12^{\circ}]$ (although the 2PACF curves show only the interval of interest: $[1.75^{\circ}, 6.5^{\circ}]$). In Fig.~\ref{fig1} we show the 2PACF curves for the six redshift {shells}, considering the cases where the number of angular bins is $N_b = 31,\, 37$, and $40$. In each case the bin size is given by $\Delta \theta = 12^{\circ} / N_b$.

As an example, one can see in Fig.~\ref{fig1} that for $\bar{z}=0.55$, the first bump located in 
$\theta \in [3^{\circ},  4^{\circ}]$ is present 
for the analysis considering 37 and 40 bins but its amplitude is significantly reduced ($\sim$ 30 \%) when 31 bins are considered. For comparison, the amplitude of the second bump, $\theta \in [4^\circ, 5^\circ]$, decreases only by 
$\sim$ 6 \% in these analyses, which shows its robustness relative to the first one, a clear indication that it 
corresponds to the acoustic scale. 
Similar results were also found for the other redshift shells.

\begin{figure*}[t]
\includegraphics[width=16.cm,height=11.5cm]%[scale=0.7]
{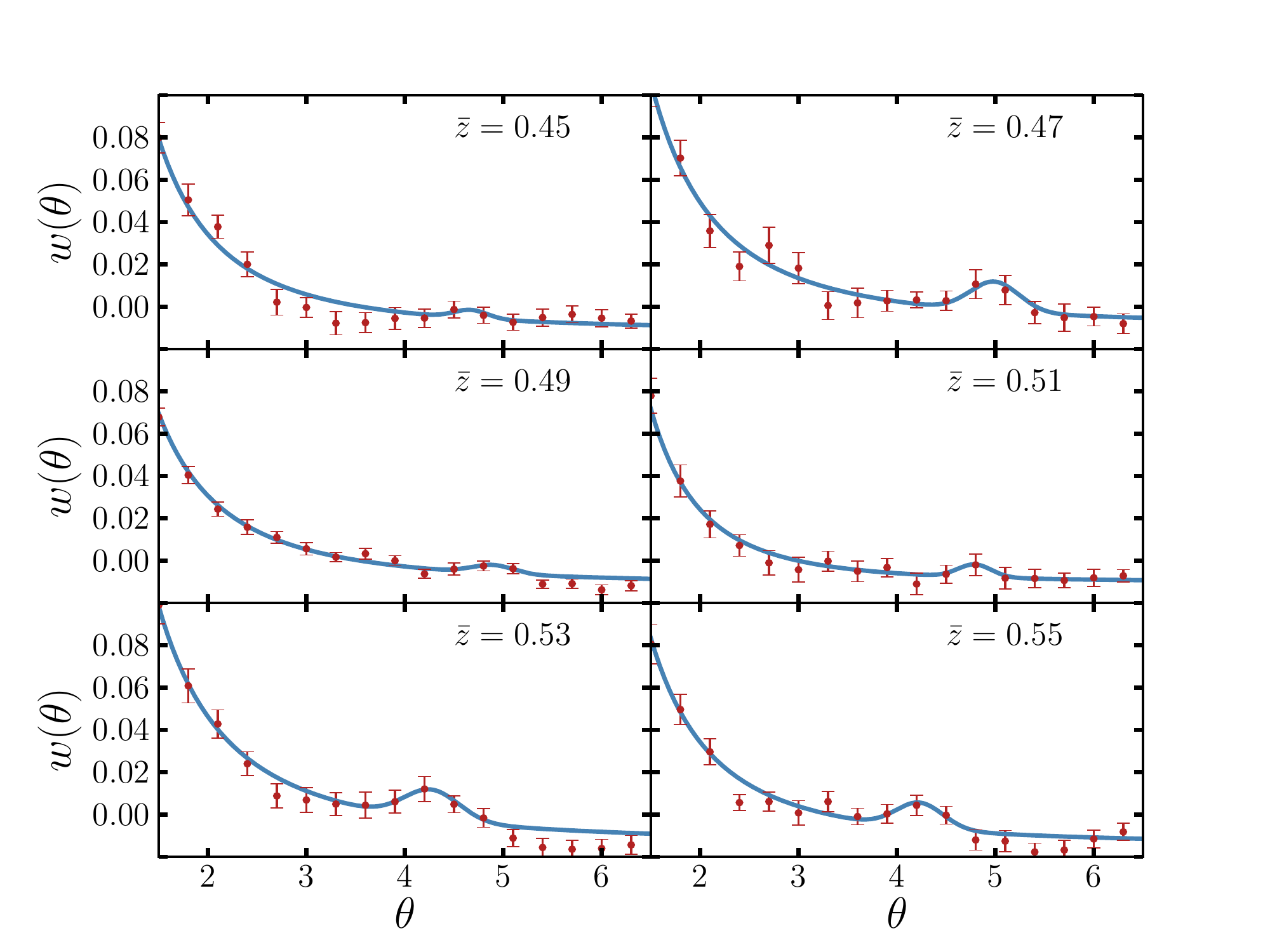}
\caption{The 2PACF for six bin redshift intervals using the DR10-SDSS data (bullets) and Eq.~\ref{eq:fit} 
(continuous line). 
The amplitude of the BAO bump corresponds to $C$, the BAO location and the width are related 
to $\theta_{FIT}$ and $\sigma$, respectively. In these plots we used $N_b = 40$.} 
\label{fig3}
\end{figure*}

\subsubsection{Small shifts of the galaxies angular coordinates}

Various peaks or bumps can be observed in the 2PACF curves shown in Fig.~\ref{fig1}. {They contain information regarding not only the true BAO signature}, but also to systematic effects present in the distinct redshift samples used in our analysis. Distinguishing the BAO bump from systematic bumps is the aim of the algorithm presented bellow. The procedure is based on the hypothesis that the primordial BAO signature is present in the catalog. According to this, we argue that if such signature is present, then it manifests itself as a robust bump at a given angular scale $\theta_{BAO}$. Instead, systematic effects, like signals produced by groups or clusters of galaxies, contribute with bumps at several angular scales, and do not survive to small perturbations of the galaxy positions. Consequently, one expects they will make no contribution to the 2PACF.

Our algorithm follows two steps. First, we derive the 2PACF corresponding to the case where no BAO signature is present at all. This function is obtained averaging one hundred 2PACFs, each one obtained by changing the angular positions of the galaxies by a random amount. The random displacements were performed following Gaussian distributions with $\sigma=0.25,0.5,$ and $1.0$, which are equivalent to maximum displacements of $\theta \sim 1.25^\circ, 2.5^\circ$, and $5^\circ$, respectively.
In principle, this procedure can destroy any bump, resulting in a smooth averaged curve, as observed in the panels 
of Fig.~\ref{fig2} (although similar results were obtained with $N_b=31$ and $37$ bins, we used 40 bins in the  analyses shown in Figs.~\ref{fig2} and~\ref{fig3}). In fact, one notices that for the cases with large values of $\sigma$, e.g., $\sigma =0.5$ and 1, all the peaks disappear. On the other hand, for $\sigma=0.25$ only one peak remains, showing the robustness of the feature expected for the BAO signature. 
The second step is to compare the original 2PACF with the smoothed no-BAO curve corresponding to the case $\sigma= 0.5$, case in which the noise is fully removed and comparison with the original 2PACF allows to identify the position of the BAO signal (the bin size criterium is also adopted to confirm this result). We then identify the BAO signature as corresponding to the excess in the original 2PACF with respect to the smoothed no-BAO curve, 
features that are more easily seen, e.g., in the shells $\bar{z}=0.47$ and $\bar{z}=0.55$. 
We emphasize that both criteria are applied to all redshift shells.

\subsection{Obtaining the $\theta_{FIT}$ values} 

After {finding} the real BAO signature, we obtain the angular BAO scale using the method of Ref.~\cite{Sanchez11}, which {parameterises} the 2PACF as a sum of a power law, describing the continuum, and a Gaussian peak, which describes the BAO bump, i.e.,
\begin{equation} 
w_{FIT}(\theta) = A  + B \theta^\nu + C e^{-\frac{(\theta-\theta_{FIT})^2}{2 \sigma_{FIT}^2}}  \, ,
\label{eq:fit}
\end{equation}
where $A,B,C,\nu$, and $\sigma_{FIT}$ are free parameters, $\theta_{FIT}$ defines the position of the acoustic scale and $\sigma_{FIT}$ gives a measure of 
the width of the bump. If $\delta z = 0$, the true BAO scale $\theta_{BAO}$ and $\theta_{FIT}$ would coincide. However, for $\delta z \neq 0$ this is no longer true because of projection effects due to the width of the redshift {shells}. Therefore, the angular correlation function given by Eq.~(\ref{expected}) has to be calculated for both $\delta z = 0$ and $\delta z \neq 0$ so that one can compare the position of the peak in the two cases. This will allows one to find a correction factor $\alpha$ that, given the value of $\theta_{FIT}$ found using relation (\ref{eq:fit}), will provide the value for $\theta_{BAO}$.

\begin{table}[t]
 \begin{tabular}{|  l | c| c | c | c | c| }
  \hline
 \,\,Models & \,\,\,\,$\Omega_b h^2$\,\,\,\, & \,\,\,\,$\Omega_c h^2$\,\,\,\, & \,\,${\rm{w}}_0$\,\, 
 & \,\,${\rm{w}}_a$\,\,& \,\,$H_0$\footnote{in units of km/s/Mpc} \,\, \\%(Km/s/Mpc)\,\,  \\
   \hline \hline
 \,\,Reference\,\,         & \,\,\,\,0.0226\,\,\,\, &\,\,\,\,0.112\,\,\,\, & \,\,-1 &  \,\,0 & 70 \\% reference model
 \hline \hline 
 \,\,Varying $\Omega_ch^2$\,\,&\,\,\,\,0.0226\,\,\,\,& \,\,\,\,0.100\,\,\,\, & \,\,-1 & \,\,0 &  70\\% Omc007
                         & \,\,\,\,0.0226\,\,\,\, & \,\,\,\,0.140\,\,\,\, & \,\,-1 & \,\,0 & 70 \\% Omc017
   \hline \hline
 \,\,Varying state     & \,\,\,\,0.0226\,\,\,\, & \,\,\,\,0.112\,\,\,\, & \,\,-2 & \,\,0 & 70 \\ %w0-2
  \,\,equation\,\,       & \,\,\,\,0.0226\,\,\,\, & \,\,\,\,0.112\,\,\,\, & \,\,-0.8\,\, & \,\,0 & 70 \\ %w0-08  
                & \,\,\,\,0.0226\,\,\,\, & \,\,\,\,0.112\,\,\,\, & \,\,-1 & \,\,1 & 70 \\ %wa1  
                & \,\,\,\,0.0226\,\,\,\, & \,\,\,\,0.112\,\,\,\, & \,\,-1 & \,-1 & 70 \\ %wa-1  
   \hline \hline      
 \,\,Varying $H_0$\,\,         & \,\,\,\,0.0226\,\,\,\, & \,\,\,\,0.112\,\,\,\, & \,\,-1 &  \,\,0 & 65 \\ %H_3
                  & \,\,\,\,0.0226\,\,\,\, & \,\,\,\,0.112\,\,\,\,   & \,\,-1 &  \,\,0 & 68 \\ %H_1
                   & \,\,\,\,0.0226\,\,\,\, & \,\,\,\,0.112\,\,\,\,  & \,\,-1 &  \,\,0 & 72 \\ %H_2
                   & \,\,\,\,0.0226\,\,\,\, & \,\,\,\,0.112\,\,\,\,  & \,\,-1 &  \,\,0 & 75 \\ %H_4 
   \hline
\end{tabular}
\caption{Summary of the cosmological models parameters considered in our analysis.}
\label{tab2}
\end{table}

\begin{table*}[]
\begin{tabular}{|  l | c | c | c | c | c | c |} 
\hline
\,\,\,\,\,\,$z$  interval\,\,\,\,\, & \,\,$\bar{z}$\,\, & \,\,$\alpha$\, (\%) \,\, 
& \,\,\,$\theta_{FIT}$ ($^{\circ}$) \,\,\, 
& \,\,$\theta^{0}_E(\bar{z})$ ($^{\circ}$)
&  \,\,$\theta_{BAO}$ ($^{\circ}$) \,\, & \,\,$\sigma_{BAO}$\,\, 
%& $D_A^{\Lambda CDM}(z)$ 
\,\, \\
\hline \hline
\,\,\,0.440-0.460 & 0.45 & 2.0815 &  4.67&  4.96 & 4.77
& 0.17  \\
\,\,\,0.465-0.475 & 0.47 & 0.5367&  4.99&  4.77 & 5.02
& 0.25  \\
\,\,\,0.480-0.500 & 0.49 & 2.0197 &  4.89 &  4.60 & 4.99
& 0.21  \\
\,\,\,0.505-0.515 & 0.51 & 0.5002 &  4.79 &  4.44 & 4.81
& 0.17  \\
\,\,\,0.525-0.535 & 0.53 & 0.4847 &  4.27 &  4.29 & 4.29
& 0.30  \\
\,\,\,0.545-0.555 & 0.55 & 0.4789 &  4.23 &  4.16 & 4.25
& 0.25  \\
 \hline
\end{tabular}
\caption{Estimates of $\theta_{BAO}(z)$ from SDSS DR10 LRG data.} 
\label{tab3}
\end{table*}

In order to calculate the 2PCF given by Eq. (\ref{eq:xiexp}), one needs the theoretical mater power spectrum, $P_m(k,z)$. We use the CAMB software (Code for Anisotropies in the Microwave Background)~\cite{camb}\footnote{http://www.camb.info}, and assume a varying dark energy model with w$(a)$ = $\rm{w}_0$ + w$_a(1-a)$, where $a$ is the cosmological scale factor. In the present analysis we assume a minimal model using six cosmological parameters 
\begin{equation}\label{parameter} 
\{ \omega_b,\omega_c, \Theta,\tau,\mathcal A_s, n_s \} \, ,
\end{equation}
where $\omega_b = \Omega_b h^2$ and $\omega_c = \Omega_c h^2$ are, respectively, the baryon and cold dark matter densities, $\Theta$ is the ratio between the sound horizon and the angular diameter distance at decoupling, $\tau$ is the optical depth to reionization, $\mathcal A_s$ is the overall normalization of the primordial power spectrum, and $n_s$ is the effective tilt. 
We consider purely adiabatic initial conditions, impose flatness and set neutrino masses equal to $3.046 \,eV$. We set the parameter values of the {reference} cosmological model 
used in this work as follows: 
$\omega_b h^2= 0.0226,\, \omega_c h^2=0.112,\, 100\, \Theta=1.04,\, \tau=0.09,\, \mathcal A_s e^9 = 2.2,\, n_s= 0.96$ with $H_0 = 100 h$ km/s/Mpc. 

For each value  $z=\bar{z}_i$, the power spectrum $P(k, z)$ was calculated  and tabulated for values of $k$ in the range $[\, 10^{-4} \,-\, 2.38 \,]\, h$ Mpc$^{-1}$. From this table, we numerically calculate the integral in Eq. (\ref{xi_e}) to obtain the spatial correlation function $\xi_E(s;\bar{z}_i)$. In order to minimize numerical uncertainties, we integrate  analytically between neighboring points of the table by drawing straight lines. The fact that, in performing the numerical integration, we can not extend the upper limit of the integral to infinity, gives rise to small oscillations that we smoothed out using a standard moving average filter. The integral which gives the angular correlation function (Eq. \ref{expected}) was calculated using a top-hat distribution for the redshift shell selection function $\phi$. We then numerically integrate $w_{E}(\theta)$ for the redshift {shells} $\bar z$ and widths $\delta z$ listed in Table I and found the position of the peak $\theta_{E}^{\delta z}
$. By similarly proceeding and putting $\delta z = 0$, we obtain $\theta_{E}^{0}$ and calculate the shift factor, defined as
\begin{equation}
\alpha={({\theta_{E}^{0} - \theta_{E}^{\delta z}}) \over {\theta_{E}^{0}}}\;.
\end{equation}
Thus, the $\theta_{BAO}$ values will be given by the expression
\begin{equation} \label{alpha}
\theta_{BAO}(z, \delta z) \,=\, \theta_{FIT}(z) + \alpha(z, \delta z)\theta_{E}^{0}(z) \, .
\end{equation}
We calculated the shift factor $\alpha$ for several cosmologies (see Table II) in order to study its dependence upon the cosmological paremeters. Our overall conclusion is that the shift factor $\alpha$ is almost independent of the cosmological parameters in the range of values here considered, with the difference between $\theta_{FIT}$ and $\theta_{BAO}$ being $\lesssim 2\%$.  We also noticed that the small value of $\alpha$ is mainly due to the tiny redshift {shell}  $\delta z \le 0.02$, chosen in our analyses, in that the lesser the value of $\delta z$, the lesser the shift factor $\alpha$ is (for comparison, we refer the reader to Fig. 3 of \cite{Sanchez11}). The results for the reference model are shown in Table III.

\begin{figure*}[t]
\mbox{\hspace{-0.8cm}
\includegraphics[width = 7.6in, height = 2.2in]{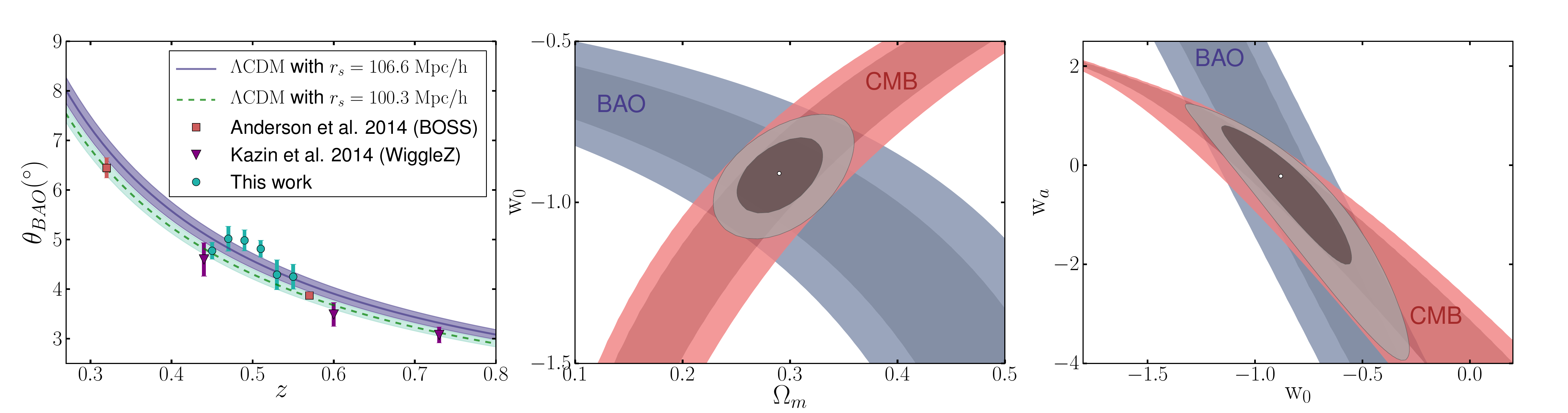}}
\caption{{\it{Left)}} The angular BAO scale as a function of redshift.  As indicated in the figure, the blue data points correspond to the 
six measurements obtained in this paper (Table III) whereas the curves stand for the $\Lambda$CDM prediction with the acoustic scale fixed at the WMAP9 and Planck values. {\it{Central)}} Confidence contours in the $\Omega_m$ - w$_0$ plane. Note that the combination between $\theta_{\rm{BAO}}(z)$ and CMB sharply limits the allowed range of the cosmological parameters. {\it{Right)}} The same as in the previous panel for the w$_0$ - w$_a$ plane.} 
\label{fig4}
\end{figure*}

\begin{figure}[h]
\mbox{\hspace{-1.0cm}
\includegraphics[scale=0.4]{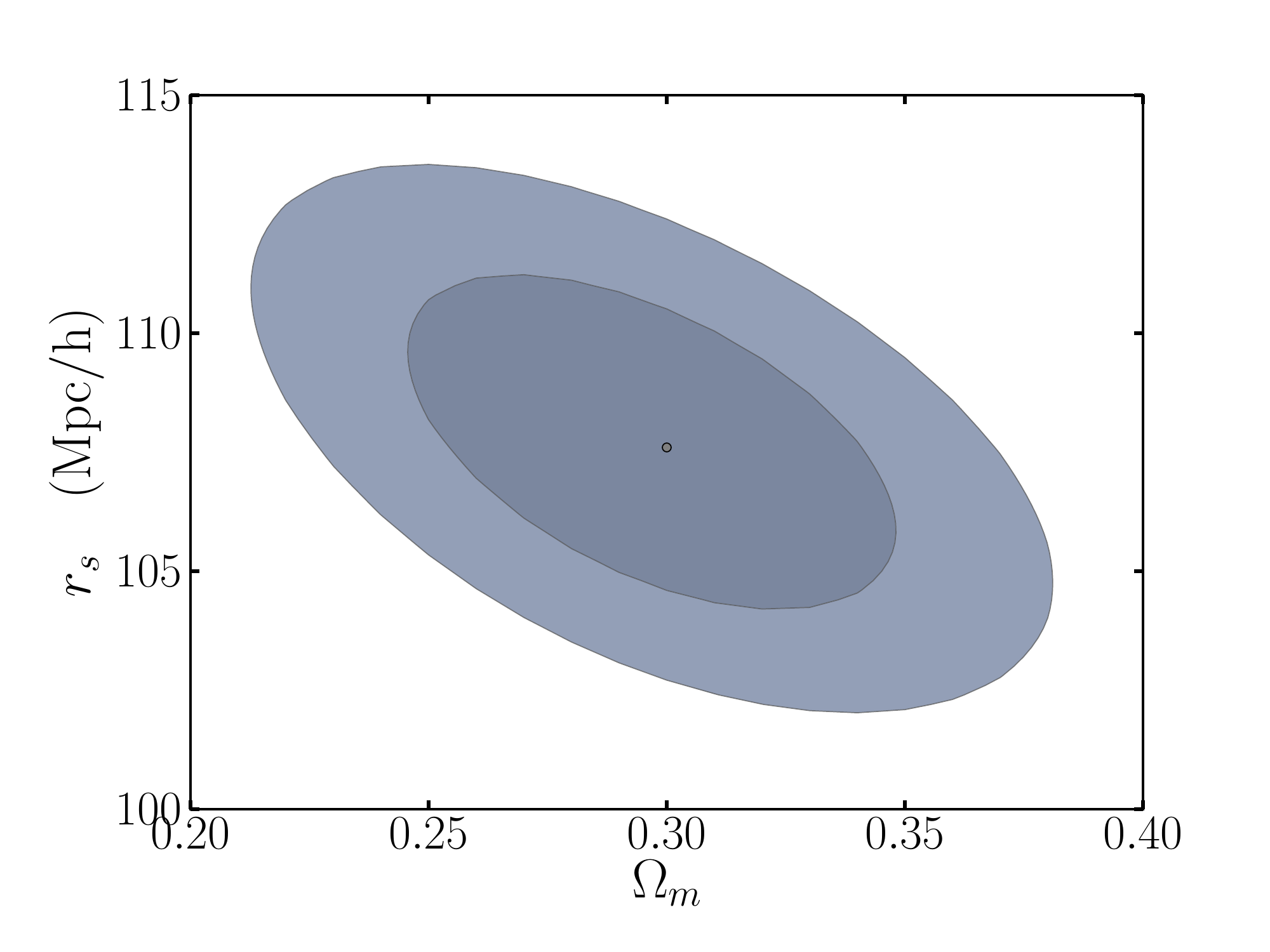}}%[scale=0.42]{rsom_prior}}
\caption{The $\Omega_m$ - $r_s$ plane obtained from the $\theta(z)$ data displayed in Table III assuming the estimate of the matter density parameter from current type Ia supernova data~\cite{sne}.} 
\label{fig4}
\end{figure}

\section{\label{CosMod} Cosmological constraints}

In this Section we present cosmological parameter fits to the BAO data displayed in Table III. The angular scale $\theta_{BAO}$ is related to the angular diameter distance $D_A(z)$ {through}
\begin{equation} \label{thetaz}
 \theta_{BAO} (z)  = \frac{r_s }{(1+ z) D_A(z)}\,,
\end{equation}
where $D_A(z) = r(z)/(1+z)$ and $r(z)$ is given by Eq. (\ref{rz}). The evolution of $\theta_{BAO}$ with redshift, $z$, is presented in Figure 4 (left panel). The lines and coloured bands correspond to the standard $\Lambda$CDM cosmology, assuming $\Omega_m=0.27$, for two values of the comoving  acoustic scale, i.e., $r_s = 106.61 \pm 3.47$ $h^{-1} \rm{Mpc}$ and $r_s = 100.29 \pm 2.26$ $h^{-1} \rm{Mpc}$ given by the WMAP9~\cite{wmap9} and Planck data~\cite{planck}, respectively\footnote{ An almost model-independent measurement of the BAO ruler length was recently reported in Ref.~\cite{Heavens}. Using type Ia supernova and galaxy clustering data, the authors obtained $r_s = 101.9 \pm 1.9 h^{-1}{\rm{Mpc}}$.}. Along with the six data points obtained in our analysis, we also show other estimates obtained by translating the three-dimensional averaged distance parameter $D_{\rm{v}}(z) = [(1+z)^2D_A^2(z)cz/H(z)]^{1/3}$ into an angular scale, with the fiducial cosmology used in each reference. Such estimates are 
shown only for 
comparison  and will not be used in the statistical analysis that follows.

As mentioned earlier, we consider a varying dark energy model whose equation-of-state parameter evolves linearly with the scale factor, w$(a)$ = $\rm{w}_0$ + w$_a(1-a)$ (w$(a)$CDM) and some of its particular cases, i.e., w$_0 \neq -1$ and w$_a = 0$ (wCDM) and w$_0 = -1$ and w$_a = 0$ ($\Lambda$CDM). Plots of the resulting cosmological constraints are shown in Figure 4 (central and right panels).  We find a tension between the $\theta_{BAO}(z)$ data derived in this analysis and the value of the acoustic scale $r_s$ given by the Planck collaboration (see left panel of Fig. 4). For instance, assuming the wCDM cosmology, the combination of both data provides $\Omega_m= 0.39\pm 0.03$ and w$_0 = -0.68 \pm 0.075$. 

In what follows, we only display the results obtained assuming the acoustic scale estimate from WMAP9. Clearly, the $\theta_{BAO}$ data alone (blue contours) are consistent with a wide range of w$_0$ and w$_a$ values, with the best-fit values being compatible with phantom scenarios in which w $< -1$. On the other hand, the combination with the CMB data sharply limits the allowed range of w, favouring values of w $ \simeq -1.0$. This can be seen when we combine the BAO data points with measurements of the shift parameter (red contours), defined as ${\cal{R}} = \sqrt{\Omega_m}\int^{z_{ls}}_{0}H_0/H(z) dz$, where $z_{ls}$ is the redshift of the last scattering surface. To be consistent with the $r_s$ value used in the BAO analysis, we use ${\cal{R}} = 1.728 \pm 0.016$, also determined with model parameters set by WMAP9 data~\cite{wmap9}. The joint results (grey contours) improve significantly the cosmological constraints, providing $\Omega_m = 0.29 \pm 0.02$ and w$_0 = -0.91 \pm 0.08$ (wCDM) and w$_0 = -0.88 \
pm 0.20$ and w$_a = -0.22 \pm 0.9$ (w$(z)$CDM) at 68.3\% C.L. Assuming  w $= -1$ ($\Lambda$CDM), we find $\Omega_m =0.334 \pm 0.054$ also at 68.3\% C.L.

%These and the results obtained assuming the comoving acoustic scale from the Planck data are shown in Table IV. In this latter case, the value of the shift parameter used in the joint analysis is the one reported in Ref.~\cite{wang}, i.e., ${\cal{R}} = 1.7407 \pm 0.0094$.

Finally, it is important to mention the role of the acoustic scale in the $\theta_{BAO}$ analysis [see Eq. (\ref{thetaz})]. As discussed above, most of the estimates of this quantity currently available were obtained from CMB data (an exception being the measurement reported in Ref.~\cite{Heavens}). From Eq. (\ref{thetaz}), however, we can directly estimate  $r_s$ from the $\theta_{BAO}$ data displayed in Table III, assuming a given cosmology. For the $\Lambda$CDM scenario, we find $r_s = 101.2 \pm 11.8$ $h^{-1} \rm{Mpc}$, which is in good agreement with both the WMAP9 and Planck values as well as with value obtained in~\cite{Heavens}. In Fig. 5 we show the $\Omega_m$ - $r_s$ plane obtained when one assumes $\Omega_m = 0.295 \pm 0.034$, as derived from current type Ia supernova data~\cite{sne}. In this case, we find $r_s= 107.6 \pm 2.3$ $h^{-1} \rm{Mpc}$, which is in full agreement with the $\Lambda$CDM estimate of $r_s$ provided by the WMAP9 analysis and $\sim 1.5\sigma$ off from the 
value of the Planck collaboration.

\section{Conclusions}

The baryon acoustic oscillations signal imprinted in the galaxy distribution is a key prediction of cosmological models, 
depending on the sound speed and expansion rate during decoupling. 
A decade after the first BAO detections, measurements of the BAO scale has become one of the main tools of precision 
cosmology which can be used to place sharp constraints on the main cosmological parameters, 
 using the data from future large-volume galaxy surveys like, e.g., JPAS~\cite{JPAS}.

We have analysed the 2PACF of luminous red galaxies from the SDSS-DR10  data and measured the BAO signal in the distribution of galaxies at six redshift {shells} in the interval $0.440 \le z \le 0.555$. Differently from the 2PCF analysis, the use of 2PACF involves only the angular separation between pairs, yielding model-independent information about $\theta_{\rm{BAO}}(z)$. In practice, however, at least two relevant sources of model-dependence can be identified. First, due to projection effects ($\delta z \neq 0$), the true BAO scale $\theta_{BAO}$ differs from  $\theta_{FIT}$, with the shift factor $\alpha$ between them depending on the predicted power spectrum $P(k,z)$ of a given cosmology. For the range of cosmological parameters displayed in Table II and the width of redshift shell considered in our analysis ($\delta z \leq 0.02$), we have found $\alpha \lesssim 2\%$. 
Nevertheless, as observed in Table~\ref{tab3}, the largest shift obtained considering several cosmologies is $0.1^{\circ}$, which is one-third of the 
size of bin $\Delta \theta$ in the 2PACF curves. In other words, this amounts to say that, in practical terms, the model-dependent shift is negligible. 
Second, in order to estimate the cosmological parameters from Eq. (\ref{thetaz}), an independent measurement of the acoustic scale is 
needed. However, as discussed in Sec. V, the current estimates of  $r_s$ are not completely model-independent, with most of them being obtained by setting the $\Lambda$CDM parameters extracted from the CMB data. 

We have also introduced and applied a model-independent methodology to identify the true BAO bump. Such a procedure is different from what has usually been done in the previous literature in which a given cosmology is taken as a guide to identify the true acoustic scale. After applying this methodology, we have derived a new  $\theta_{\rm{BAO}}(z)$ sample, which increases significantly the current number of $\theta_{\rm{BAO}}(z)$ data points available in the literature. From this $\theta_{\rm{BAO}}(z)$ sample, we have derived cosmological constraints on a class of dark energy scenarios with w$(a)$ = $\rm{w}_0$ + w$_a (1-a)$. As shown in Fig. 4 (central and right panels), although the $\theta_{\rm{BAO}}(z)$ data alone are consistent with a wide range of w$_0$ and w$_a$ values, the combination with the CMB data sharply constrains the allowed range of the dark energy equation of state, favoring values of w compatible 
with a cosmological constant. 

A final aspect worth mentioning is the possibility of use the current $\theta_{\rm{BAO}}(z)$ sample to estimate some relevant cosmological parameters such as the present expansion rate, $H_0$, and the number and mass of neutrino species. In principle, this can be done by calculating the shift correction $\alpha$, defined in Eq. (\ref{alpha}), for a given cosmology. A detailed study exploring this possibility is currently in progress and will appear in a forthcoming communication.

\section*{Acknowledgements}

The authors thank CNPq, CAPES, INEspa\c{c}o and FAPERJ for the grants under which this work was carried out. Joel C. Carvalho is also supported by the DTI-PCI  program of the Brazilian Ministry of Science, Technology and Innovation (MCTI).

%\newpage

\end{document}